\documentclass[12pt,a4paper]{article}
\usepackage{epsfig,amssymb}

\begin{document}
\title{
Physical conditions in potential sources of ultra-high-energy cosmic rays:
  Updated Hillas plot and radiation-loss constraints
}
\author{
Ksenia Ptitsyna$^1$
and Sergey Troitsky$^2$\footnote{st@ms2.inr.ac.ru}\\
$^1$~M.V.~Lomonosov Moscow State University,
Moscow 119992, Russia\\
$^2$~Institute for Nuclear Research of the Russian Academy of
Sciences,\\
60th October Anniversary Prospect 7a, 117312, Moscow, Russia
}

\date{}
\maketitle
\abstract{
We review basic constraints on the acceleration of ultra-high-energy (UHE)
cosmic rays (CRs) in astrophysical sources, namely the geometrical
(Hillas) criterion and restrictions from radiation losses in different
acceleration regimes. Using the latest available astrophysical data, we
redraw the Hillas plot and figure out potential UHECR accelerators.
For the acceleration in central engines of
active galactic nuclei, we constrain the maximal UHECR energy for a given
black-hole mass.
Among active galaxies, only the most
powerful ones, radio galaxies and blazars, are able to accelerate
protons to UHE, though acceleration of heavier nuclei is possible in
much more abundant lower-power Seyfert galaxies.
}


\maketitle

\tableofcontents

\section{Introduction}
\label{sec:intro}

\markboth{Physical conditions in potential UHECR sources I}{
Physical conditions in potential UHECR sources I}

The origin of ultra-high-energy (UHE; energy ${\cal E} \gtrsim 10^{19}$~eV)
cosmic rays (CRs) remains unknown despite decades of intense studies (see
e.g.\ Ref.~\cite{NaganoWatson} for a comprehensive review and
Ref.~\cite{Kachelriess:lectures} for a recent pedagogical introduction).
Recent studies, notably the observation of the
Greisen--Zatsepin--Kusmin~\cite{G, ZK} cutoff by the HiRes
experiment~\cite{HiRes:cutoff}, further supported by results of the Pierre
Auger observatory (PAO)~\cite{PAOspectrum}, suggest that at least a large
fraction of UHECRs is accelerated in cosmologically distant astrophysical
sources. The birth of (scientific) UHECR astronomy, however, awaits our
firm understanding of energies and primary composition of the observed
cosmic rays as well as identification of at least one astrophysical object
where these particles are accelerated.

Given  experimental ambiguities, it is important to understand
theoretically, which astrophysical objects may serve as UHECR
accelerators. It has been understood long ago that the UHECR sources
should be distinguished objects with extreme physical conditions. One
simple criterion is the geometrical one: the particle should not leave the
accelerator before it gains the required energy. Presumably, the particle
is accelerated by the electric field and confined by the magnetic one;
then the geometrical criterion is expressed in terms of the particle's
Larmor radius which should not exceed the linear size of the accelerator
(see, e.g., Ref.~\cite{StellarMagnetospheres}). In the context of UHECRs,
this condition is recognized as the Hillas criterion~\cite{Hillas} and is
often presented graphically in terms of the Hillas plot where the
accelerator size $R$ and the magnetic field $B$ are plotted. We note that
even quite recent reviews use either cut-and-pasted or slightly refurbished
versions of the original 25-years-old plot. However, astrophysics
experienced enormous progress, if not a revolution, during these decades.
One of the aims of this study is to give an updated version of the Hillas
plot with references to either -- when possible -- measurements or
estimates of the magnetic fields and sizes of potential astrophysical
accelerators. The most important update corresponds to a wide variety of
active galaxies whose sizes and magnetic fields differ by many orders of
magnitude from one object to another so that some of them may, while most
of them may not, accelerate particles to UHE.

Another restriction on the cosmic-ray accelerators is posed by the
radiation losses which inevitably accompany the acceleration of a charged
particle. The corresponding constraints were studied, in particular, in
Refs.~\cite{Hillas, Protheroe, Derishev, Medvedev}. The radiation losses
depend on the particular field configuration and the maximal achievable
energy of a particle in the loss-limited regime depends on the
acceleration mechanism. Restricting to particular mechanisms or particular
field configurations may result in would-be contradictory results, cf.\
Refs.~\cite{Derishev, Medvedev}.
In this work, we review the radiation-loss constraint for
different cases; they
further limit the acceptable region on the
updated Hillas plot. One of possible applications of these general
constraints is a study of active galaxies correlated with the Auger
events~\cite{paper2}.

One should always keep in mind that even if both geometric and radiation
constraints are satisfied, they do not yet guarantee particle acceleration
to the corresponding energy. Each particular source should be discussed in
the context of an acceleration mechanism operating there.

The rest of the paper is organised as follows. In
Sec.~\ref{sec:accel:general}, we review constraints on potential UHE
accelerators, that is model-independent Hillas geometrical constraint and
limitations due to radiation losses for particular acceleration
mechanisms. In Sec.~\ref{sec:accel:mf-measurements}, we take advantage of
the modern astrophysical data and redraw the Hillas plot supplemented by
the radiation-loss constraints. Our results are summarized  and
discussed in Sec.~\ref{sec:accel:summary} while brief conclusions are
given in Sec.~\ref{sec:concl}. Appendix~\ref{app:ed} contains derivation
of some formulae.

\section{General constraints from geometry and radiation}
\label{sec:accel:general}

An accelerator of UHECR particles should satisfy several general
constraints which may be briefly summarized as follows:
\begin{itemize}
 \item
{\bf geometry} --- the accelerated particle should be kept inside the
source while being accelerated;
\item
{\bf power} --- the source should possess the required amount of energy to
give it to accelerated particles;
\item
{\bf radiation losses} ---
the energy lost by a particle for radiation in
the accelerating field should not exceed the energy gain;
\item
{\bf interaction losses} ---
the energy lost by a particle in interactions with other particles
should not exceed the energy gain;
\item
{\bf emissivity} ---
the total number (density) and power of sources should be able to provide
the observed UHECR flux;
\item
{\bf accompanying radiation} of photons, neutrinos and low-energy cosmic
rays should not exceed the observed fluxes, both for a given source and
for the diffuse background (in particular, the ensemble of sources should
reproduce the observed cosmic-ray {\em spectrum}\footnote{We note that
the spectrum of cosmic rays accelerated in a particular source may be
very different from the spectrum observed at the Earth, cf.\
Ref.~\cite{Semikoz:spectrum}.}).
\end{itemize}
The primary concern of this paper is the geometrical and radiation-loss
constraints (others are briefly quoted when relevant); both restrict the
magnetic field and the size of the accelerator and can be graphically
represented on the Hillas plot.

\subsection{The Hillas criterion}
\label{sec:accel:general:Hillas}
The Larmor radius $R_L$ of a particle does not exceed the accelerator size,
otherwise the particle escapes the accelerator and cannot gain energy
further. This Hillas criterion sets the limit
\begin{equation}
{\cal E} \le {\cal E}_{\rm H}=qBR
\label{eq:Hillas}
\end{equation}
for the energy ${\cal E}$ gained by a particle with charge $q$ in the
region of size $R$ with the magnetic field $B$.

\subsection{Radiation losses}
\label{sec:accel:general:radiation}
While Eq.~(\ref{eq:Hillas}) is a necessary limit, more stringent ones may
arise from the energy losses: the maximal energy ${\cal E}_{\rm loss}$ a
particle can get in an accelerator of infinite size is determined by the
condition
\begin{equation}
\frac{d{\cal E}^{(+)}}{dt}
=
-\frac{d{\cal E}^{(-)}}{dt},
\label{Eq:+-}
\end{equation}
where the energy gain rate in the effective electric field $E=\eta B$ is
(in the particle-physics $c=1$ units which we use throughout the paper)
\begin{equation}
\frac{d{\cal E}^{(+)}}{dt}
= q \eta B
\label{Eq:dE+}
\end{equation}
(kept explicit in equations, the efficiency coefficient $\eta$ is set to
one in figures to obtain conservative (optimistic) limits for a given
magnetic field -- the electric fields in astrophysical objects are much
less studied observationally compared to the magnetic ones, but it is
always expected that $E \ll B$). Depending on particular conditions in the
accelerator, the maximal energy ${\cal E}_{\rm max}$ of a particle is
limited either by geometrical or by energy-loss arguments:
\[
{\cal E}_{\rm max}=\min \left\{ {\cal E}_{\rm H}, \, {\cal E}_{\rm loss}
\right\} .
\]

The general expression for total radiation losses for a particle with
velocity {\bf v} moving in arbitrary electric {\bf E} and
magnetic {\bf B} fields reads (see e.g.\ Ref.~\cite{LL})
\begin{equation}
-\frac{d{\cal E}^{(-)}}{dt}
=\frac{2}{3} \frac{q^4}{m^4} {\cal E}^2 \left( \left( {\bf E} + \left[
{\bf v} \times {\bf B} \right]   \right)^2- \left( {\bf E}\cdot {\bf v}
\right)^2 \right),
\label{LL}
\end{equation}
where $q$ and $m$ are the particle's charge and mass; cross and dot
denote vector and scalar product, respectively. By making use of
relativistic equations of motion, Eq.~(\ref{LL}) can be conveniently
rewritten~\cite{Longair} as
\[
-\frac{d{\cal E}^{(-)}}{dt}
=\frac{2}{3}
\frac{q^2}{m^2(1-v^2)}
\left({\bf F}^2 - \left({\bf F}\cdot {\bf v} \right)^2  \right).
\]
The force ${\bf F}$ acting on a particle is further decomposed as ${\bf
F}={\bf F}_\parallel +{\bf F}_\perp$, where we determine the parallel
${\bf F}_\parallel$ and perpendicular ${\bf F}_\perp$ components with
respect to ${\bf v}$, that is $({\bf F_\perp}\cdot {\bf v})=0$. Then
\begin{equation}
-\frac{d{\cal E}^{(-)}}{dt}
=\frac{2}{3}
\frac{q^2}{m^2(1-v^2)}
\left(F_\perp^2 + F_\parallel^2 (1-v^2)  \right).
\label{5*}
\end{equation}
It is apparent that the second term (the so-called curvature radiation)
is suppressed with respect to the first one (synchrotron radiation) by an
extra power of $(1-v^2) $ and therefore may be neglected in the
ultrarelativistic regime unless the synchrotron term is zero or very small
itself. The synchrotron losses are dominant for any generic field
configuration; however, in a very specific regime ${\bf v} \parallel {\bf
E} \parallel {\bf B}$ they vanish, and the losses are then determined by
the curvature radiation.

\subsection{Different acceleration regimes}
\label{sec:accel:regimes}
Depending on the scenario of acceleration, we will consider diffusive
(stochastic) and inductive (one-shot, or direct) mechanisms (see e.g.\
Ref.~\cite{Hillas} for a general discussion of these two approaches to
UHECR acceleration).

The prime examples of diffusive processes are the Fermi first-order
\cite{Fermi} and second-order (shock, e.g.~\cite{BlandEich}) acceleration.
Other possibilities include interaction with medium by crossing a boundary
between layers with different velocities \cite{OstrLayers} and even
transformation of a particle into a different one \cite{DerishevStern}. A
recent review and more references can be found in Ref.~\cite{Ostrowski08}.

In inductive mechanisms,
the particle is accelerated by the large-scale electric field continuously
and then leaves the accelerator. Strong ordered fields on relatively large
scales are required; example scenarios are given e.g.\ in
Refs.~\cite{0106530, PSR, Neronov0, Neronov1, Neronov2}.
For our purposes, it is convenient to
separate the inductive-acceleration scenarios into two groups, depending
on whether the configuration of the accelerating field corresponds to
synchrotron- (e.g.\ large-scale jets~\cite{0106530}) or
curvature-dominated (neutron stars~\cite{PSR} and black
holes~\cite{Neronov0, Neronov1, Neronov2}) losses.

\subsubsection{Diffusive acceleration.}
\label{sec:accel:regimes:diffusive}
The losses in this regime are the most serious. This scenario cannot be
realized in strongly ordered field configurations with ${\bf v}\parallel
{\bf E} \parallel {\bf B}$, therefore the
losses are determined by the synchrotron limit,
\begin{equation}
-\frac{d{\cal E}^{(-)}}{dt}
=\frac{2}{3}\frac{q^2}{R_L^2} \left(\frac{\cal E}{m} \right)^4
=\frac{2}{3}\frac{q^4}{m^4}{\cal E}^2 B^2 ~~~~~~~~ \mbox{(synchrotron)}.
\label{Eq:loss:sync}
\end{equation}
This regime has been studied in Ref.~\cite{Medvedev} where it has been
shown (see Appendix~\ref{app:medv}) that
the maximal energy is limited by
\begin{equation}
{\cal E}_d \simeq \frac{3}{2} \frac{m^4}{q^4} B^{-2} R^{-1}.
\label{6*}
\end{equation}
Diffusive mechanisms are quite generic and may work in every realistic
environment which can host, e.g., a shock wave. Eq.~(\ref{6*}) does not
rely on a particular acceleration mechanism and gives a (hardly reachable)
upper limit for the maximal energy.

\subsubsection{One-shot acceleration with synchrotron-dominated losses.}
\label{sec:accel:regimes:sync}
In this regime, the energy loss rate is given by Eq.~(\ref{Eq:loss:sync})
and, given Eq.~(\ref{Eq:dE+}), Eq.~(\ref{Eq:+-}) results in the maximal
energy
\begin{equation}
{\cal{E}}_s=\sqrt{\frac{3}{2}} \frac{m^2}{q^{3/2}} B^{-1/2}\eta^{1/2}.
\label{6**}
\end{equation}
This acceleration mechanism requires ordered fields throughout the
acceleration site; its practical realization for UHECR
may work in jets of powerful active galaxies~\cite{0106530}.

\subsubsection{One-shot acceleration with curvature-dominated losses.}
\label{sec:accel:regimes:curv}
The energy loss rate is determined (see Appendix~\ref{app:curv}) by
\begin{equation}
-\frac{d{\cal E}^{(-)}}{dt}
=\frac{2}{3}\frac{q^2}{r^2} \left(\frac{\cal E}{m} \right)^4
~~~~~~~~ \mbox{(curvature)},
\label{Eq:loss:curv}
\end{equation}
where $r$ is the curvature radius of the field lines which is supposed to
be of order of the accelerator size and Eq.~(\ref{Eq:+-}) results in the
maximal energy
\begin{equation}
{\cal E}_c = \left(\frac{3}{2} \right)^{1/4} \frac{m}{q^{1/4}} B^{1/4}
R^{1/2}\eta^{1/4}.
\label{6***}
\end{equation}
This mechanism requires ordered fields of very specific configurations
which, however, may be present in the immediate vicinity of neutron stars
and black holes \cite{PSR, Neronov0, Neronov1, Neronov2}.

\subsection{Summary of results for the maximal energy}
\label{sec:accel:regimes:eqns}
Let us summarize the expressions for the maximal energy ${\cal E}_{\rm
max}$ (in the comoving frame) atteinable by a nuclei with atomic number
$Z$ and mass $A$ in the accelerator of size $R$ filled with magnetic field
$B$, for different acceleration regimes:
\[
{\cal E}_{\rm max}(B,R)=
\left\{
\begin{array}{ll}
{\cal E}_{\rm H}(B,R), & B\le B_0(R);\\
{\cal E}_{\rm loss}(B,R),& B>B_0(R),
\end{array}
\right.
\]
where
\[
B_0(R)= 3.16 \times 10^{-3}~{\rm G}~~
\frac{A^{4/3}}{Z^{5/3}}
\left(\frac{R}{{\rm kpc}} \right)^{-2/3}\eta^{1/3},
\]
is determined from Eqs.~(\ref{eq:Hillas}) and (\ref{6*}), (\ref{6**}) or
(\ref{6***}) by requiring $ {\cal E}_{\rm H}(B,R) = {\cal E}_{\rm
loss}(B,R)$; the Hillas constraint is
\[
{\cal E}_{\rm H} (B,R) = 9.25 \times 10^{23} ~{\rm eV}~
Z
\left(\frac{R}{{\rm kpc}} \right)
\left(\frac{B}{{\rm G}} \right)
\]
and the radiation-loss constraints are
\[
{\cal E}_{\rm loss} (B,R) =
{\cal E}_{\rm d} (B,R) =
2.91 \times 10^{16} ~{\rm eV}~
\frac{A^4}{Z^4}
\left(\frac{R}{{\rm kpc}} \right)^{-1}
\left(\frac{B}{{\rm G}} \right)^{-2}
\]
for diffusive acceleration,
\[
{\cal E}_{\rm loss} (B,R) =
{\cal E}_{\rm s} (B,R) =
1.64 \times 10^{20} ~{\rm eV}~
\frac{A^2}{Z^{3/2}}
\left(\frac{B}{{\rm G}} \right)^{-1/2} \eta^{1/2}
\]
for inductive acceleration with synchrotron-dominated losses and
\[
{\cal E}_{\rm loss} (B,R) =
{\cal E}_{\rm c} (B,R) =
1.23 \times 10^{22} ~{\rm eV}~
\frac{A}{Z^{1/4}}
\left(\frac{R}{{\rm kpc}} \right)^{1/2}
\left(\frac{B}{{\rm G}} \right)^{1/4} \eta^{1/4}
\]
for inductive acceleration with curvature-dominated losses. Applications
to particular objects and graphical representations of the constraints
will follow in Sec.~\ref{sec:accel:summary} (see in particular
Figs.~\ref{fig:HillasCurv} -- \ref{fig:HillasFe}).

Note that the critical value $B_0(R)$, at which $ {\cal E}_{\rm H}(B,R) =
{\cal E}_{\rm loss}(B,R)$, is the same for all three acceleration regimes:
in this case, the Larmor radius $R_L$ and the size of the accelerator $R$
are equal; within our approximation they coincide also with the curvature
radius $r$ of the field lines. Therefore the diffusive acceleration regime
merges the one-shot acceleration because the
particle interacts with the shock wave only once
in this limiting case; moreover, Eqs.~(\ref{Eq:loss:sync}) and
(\ref{Eq:loss:curv}) coincide and the two regimes of inductive
acceleration result in similar losses.

\section{Magnetic fields in particular sources}
\label{sec:accel:mf-measurements}
A number of astrophysical sources have been proposed where acceleration of
cosmic rays up to the highest energies can take place (see e.g.\
Refs.~\cite{sources, comparative} for reviews and summary). In this
section, we review experimental information on their magnetic fields in
order to put them in proper places on the Hillas plot. General methods of
astrophysical magnetic-field studies are discussed e.g.\ in
Refs.~\cite{Massimo, Vallee}; however, a much wider variety of them is
used
for studies of individual sources.

\subsection{Neutron stars, pulsars and magnetars}
\label{sec:accel:mf:psr}
Neutron stars host the highest known magnetic fields in the Universe. In
particular, magnetars (including anomalous X-ray pulsars) may
possess kilometer-scale fields $B\sim10^{14}$~G and higher while normal
neutron stars have $B\sim (10^{11} \div 10^{12})$~G. Observational
evidence for these high fields is discussed e.g.\ in Sec.~6.3 of
Ref.~\cite{0804.0250}. We note also a direct (though not widely accepted)
observational method to measure $B$ in neutron stars, observation of
spectral lines giving evidence for resonant Compton scattering at
cyclotron frequency in the high-field media, see e.g.\
Ref.~\cite{Nature-423-725} for normal neutron stars and
Ref.~\cite{0610382} for anomalous X-ray pulsars.

\subsection{Active galaxies}
For the purposes of the present study, we will use a simplified
classification of active galaxies (see textbooks \cite{Carroll, Postnov}
and, for a more detailed discussion, Ref.~\cite{Veron-classif}). Clearly,
there are many intermediate states and peculiar objects which do not fit
this classification well; while they should be studied individually
if suspected to be UHECR sources, their parameters of relevance (sizes and
magnetic fields) are expected to interpolate between those of better
classified active galaxies.

{\bf Seyfert galaxies} -- spiral galaxies with bright emission-line
nuclei; radio-weak; do not possess large-scale relativistic jets; often
have starburst activity.

{\bf Radio galaxies} -- radio-loud elliptical galaxies with relativistic
jets. According to Ref.~\cite{FR}, they are classified into two luminosity
classes: FR~I (less powerful; jets brighter towards core; jets may be
curved) and FR~II (most powerful; straight jets brighter at the hot spots
at their end points).

{\bf Blazars} -- (almost) point-like objects with non-thermal spectrum;
strongly variable; similar in total power to radio galaxies; may be
associated with radio galaxies whose jets are pointed towards the
observer. They may be divided into BL Lac type objects (relatively low
power; no emission lines; possible counterparts of FR~I radio galaxies)
and optically violently variable quasars (extremely powerful; may have
emission lines; possible counterparts of FR~II).

Low-power active galaxies (Seyferts) are much more abundant than radio
galaxies and blazars.

Possible acceleration sites in active galaxies include both the central
engine (immediate vicinity of the black hole and the accretion disk) and
extended structures (jets, lobes, hot spots and jet knots). We will discuss
separately the black-hole environment and extended structures because of
very different conditions for particle acceleration. Let us note that the
term ``active galactic nuclei'' (AGN) is often used to describe a region
much larger than just the black hole and its accretion disc and often
includes inner jets (or sometimes even larger structures) which we consider
separately.

\subsubsection{Supermassive black holes and their environment.}
\label{sec:accel:mf:bh}
\begin{figure}
\begin{center}
\includegraphics [width=0.7\textwidth]{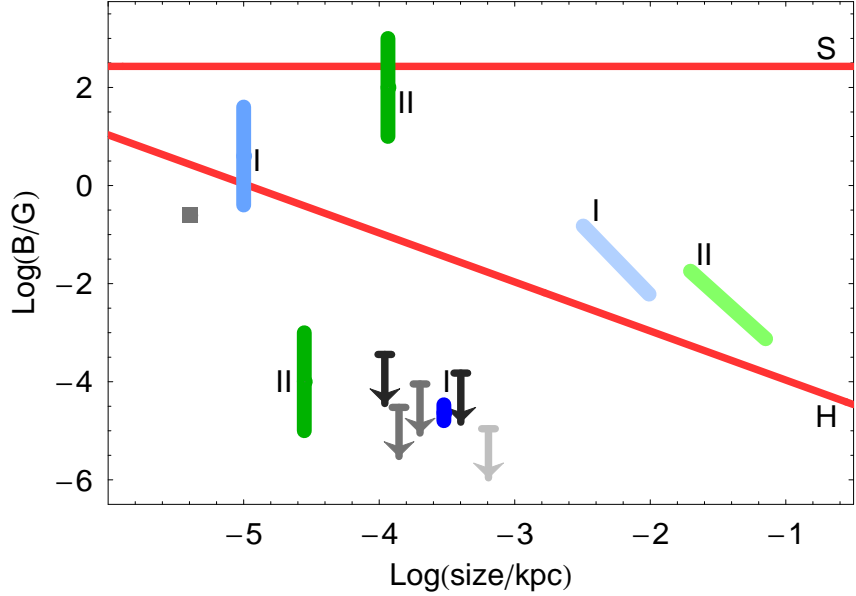}
\end{center}
\caption{
\label{fig:bh-measurements}
The size-field diagram for central regions of active galactic nuclei.
Grey colors (box and arrows) correspond to Seyfert galaxies, blue colors
(error-bar lines marked I) correspond to FRI radio galaxies, green colors
(error-bar lines marked II) correspond to FRII radio galaxies and quasars.
Arrows: upper limits from the Zeeman splitting in megamasers (light grey,
Ref.~\cite{0610912}; medium grey, Ref.~\cite{0502240}; dark grey,
Ref.~\cite{MNRAS-376-459}). Dark blue (I) vertical error bar: Faraday
rotation measurements, Ref.~\cite{ApJ-566-L9}. The grey
box~\cite{SovAL-6-42}, light blue (I) vertical~\cite{AstronRep-51-808},
dark green (II)~\cite{AstronRep-49-967} and
diagonal (I and II)~\cite{AstronRep-50-202} error bars correspond to the
measurements by the synchrotron self-absorbtion method (see text). The
allowed region for acceleration of $10^{20}$~eV protons is located between
thick red lines H and S (the lower line, H, corresponds to the Hillas
limit; the upper one, S, corresponds to the radiation-loss limit for
one-shot acceleration with synchrotron-dominated losses).}
\end{figure}

Measurements of magnetic fields in the central regions of galaxies have
been performed by means of the following methods (see
Fig.~\ref{fig:bh-measurements} for particular results).

{\bf 1. Synchrotron self-absorption.} Under certain conditions, the
low-energy cutoff in the spectrum of a compact source may be detected and
its shape may be proven to correspond to the absorption of synchrotron
photons on themselves. If this is the case, then the magnetic-field
strength may be determined by means of the Slysh formula~\cite{Slysh} or
its modifications. The method works best of all for strong radio sources
with resolved nuclear
components~\cite{SovAL-6-42, AstronRep-51-808, AstronRep-49-967,
AstronRep-50-202}.

{\bf 2. Polarimetry.} Measurements of the Faraday rotation and of
resulting depolarization give estimates of the magnetic field provided the
plasma density is known from independent observations~\cite{ApJ-566-L9}.

{\bf 3. Zeeman effect in megamasers.} Megamasers are compact sources of
coherent radiation in molecular clouds inside or around the accretion
disk. Current precision allows to put very stringent constraints on the
magnetic fields in these regions from non-observation of the Zeeman
splitting in megamasers in nearby Seyfert
galaxies~\cite{0610912, 0502240, MNRAS-376-459}.

{\bf 4. The iron $K_\alpha$ line.} Measurements of the width and shape of
this X-ray line may provide important information about circumnuclear
dynamics; in particular, it may be used to estimate the magnetic field,
though present constraints are quite weak~\cite{Lukash}.

All these direct measurements, however, cannot probe the most interesting
region in the immediate vicinity of the central black hole, a few
Schwarzschield radii ($R_S$) from the center. This region is particularly
important because theoretically motivated configurations of electric and
magnetic fields may allow for negligible synchrotron radiation of
accelerated particles and thus for (relatively weak) curvature-radiation
losses. Lack of our understanding of the field structure in the accretion
disk is transformed into uncertainties in the inferred magnetic fields
$B_{\rm BH}$ at the black-hole horizon (see e.g.\ Ref.~\cite{Gnedin} for a
summary of models used for this extrapolation). Direct estimates of
$B_{\rm BH}$ are therefore not only scarce but also model-dependent.

On the other hand, parameters of the environment of a black hole and in
particular the value of $B_{\rm BH}$ depend strongly on the black-hole
mass $M_{\rm BH}$. A conservative upper limit on $B_{\rm BH}$ follows from
the condition that the maximal rate of extraction of the rotational energy
of a black hole does not exceed the Eddington luminosity ~\cite{Znajek}
(see Ref.~\cite{Abram} for a detailed discussion),
\begin{equation}
B_{\rm BH} \lesssim 3.2 \times 10^8 \left(\frac{M_{\rm BH}}{M_\odot}
\right)^{-1/2} ~{\rm G}.
\label{ZnajekLimit}
\end{equation}
Quite old but popular models estimate the $M_{\rm BH}$ -- $B_{\rm BH}$
relation from the pressure balance (radiation pressure equals to the
magnetic-viscosity pressure)~\cite{ShakuraSyunyaev, NovikovThorne}:
\begin{equation}
B_{\rm BH} \sim 10^8 \left(\frac{M_{\rm BH}}{M_\odot}
\right)^{-1/2} ~{\rm G}.
\label{ShakuraSunyaev}
\end{equation}
An efficient method to constrain the relation between $M_{\rm BH}$ and
$B_{\rm BH}$ was found in Ref.~\cite{M-B_BH} in the frameworks of a
particular (not generally accepted) theoretical model in which both $M_{\rm
BH}$ and $B_{\rm BH}$ are related to the observable luminosity at
5100~\r{A}. It gives somewhat lower values of $B_{\rm BH}$ than
Eq.~(\ref{ShakuraSunyaev}); the best fit is
\begin{equation}
\log \left(\frac{B_{\rm BH} }{{\rm G}} \right)= \left(9.26 \pm 0.39
\right) - \left(0.81 \pm 0.05 \right) \log\left( \frac{M_{\rm BH}}{M_\odot}
 \right),
\label{M-B_BHfit}
\end{equation}
where the central values of the coefficients are taken from
Ref.~\cite{M-B_BH} and the error bars are estimated by us from their data.
For two cases when rather firm and model-independent values of $B_{\rm
BH}$ could be inferred from the observations (synchrotron self absorption
measured at different radii down to 0.1~pc and extrapolated to $R_S$,
Ref.~\cite{AstronRep-50-202}), we estimated the corresponding $M_{\rm BH}$
and found that both values are in a good agreement with
Eq.~(\ref{M-B_BHfit}), though precision is quite low.

\begin{figure}
\begin{center}
\includegraphics [width=0.7\textwidth]{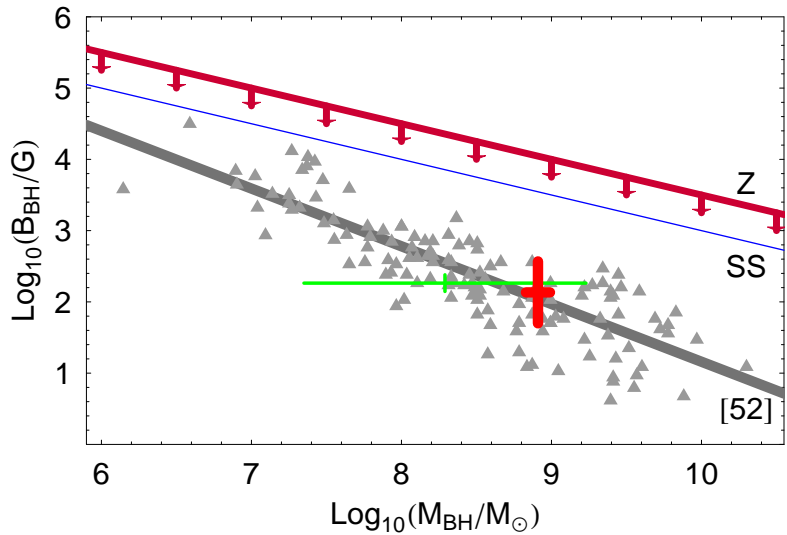}
\end{center}
\caption{
\label{fig:M-B_bh}
Magnetic field $B_{\rm BH}$ at the black-hole horizon versus the black
hole mass $M_{\rm BH}$. Triangles are
estimates of Ref.~\cite{M-B_BH} (determined in the frameworks of a
particular, not generally accepted model) and the grey line (marked
\cite{M-B_BH}) represents their best fit, Eq.~(\ref{M-B_BHfit}). Two
points
with error bars correspond to experimental estimates of $B_{\rm BH}$,
Ref.~\cite{AstronRep-50-202}, using the synchrotron self-absobtion method
($M_{\rm BH}$ estimated by us using the stellar velocity dispersion from
HyperLEDA~\cite{LEDA} (thick dark red, FRI radio galaxy 3C~465) and 2MASS
$K_s$ magnitude quoted in NED~\cite{NED} (thin light green, FRII radio
galaxy 3C~111); see Ref.~\cite{paper2} for details). Thin blue line (SS)
corresponds to the Shakura--Sunyaev estimate, Eq.~(\ref{ShakuraSunyaev}).
Thick red line (Z) represents the Znajek upper limit,
Eq.~(\ref{ZnajekLimit}). This conservative upper limit is used in our
estimates of the maximal cosmic-ray energy.}
\end{figure}

Estimates of $B_{\rm BH}$ versus $M_{\rm BH}$ are summarized in
Fig.~\ref{fig:M-B_bh}.
We will use the upper limit,
Eq.~(\ref{ZnajekLimit}), to estimate $B_{\rm BH}$ for a given $M_{\rm
BH}$; we note however that realistic values of $B_{\rm BH}$ are 1\dots 2
orders of magnitude lower. Since for the curvature-dominated radiation
losses higher $B$ always results in higher ${\cal{E}}_{\rm max}$, this
assumption is conservative for our purposes.

\begin{figure}
\begin{center}
\includegraphics [width=0.7\textwidth]{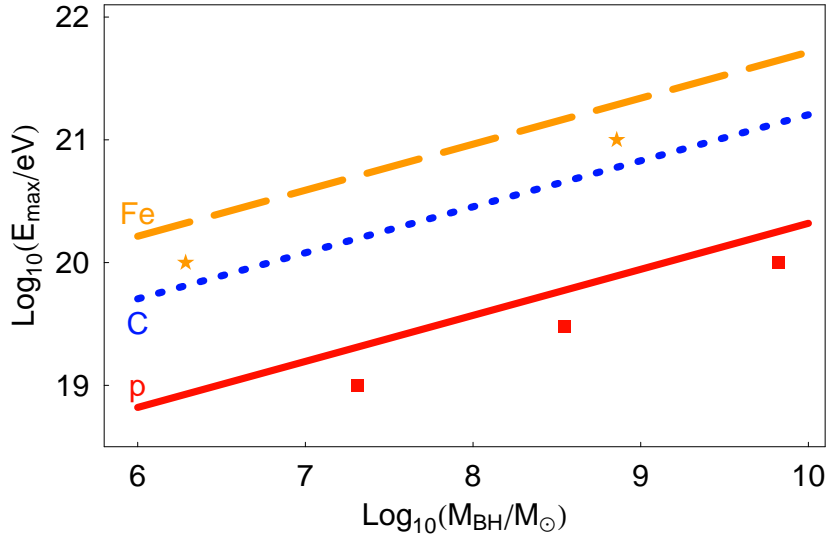}
\end{center}
\caption{
\label{fig:M-E_bh}
Upper limit on the maximal atteinable energy of protons (red solid line),
carbon nuclei (blue dotted line) and iron nuclei (orange dashed line) for
acceleration with curvature-dominated losses near a supermassive black
hole, Eq.~(\ref{Eq:Emax-MBH}). The maximal energy obtained in numerical
simulations in a particular mechanism~\cite{Neronov2} is shown by red
boxes (protons) and orange stars (iron nuclei); these data were obtained
from Figs.~5 and 10 of Ref.~\cite{Neronov2} and Eq.~(\ref{ZnajekLimit}) of
this paper.}
\end{figure}

The size $R$ of the potential acceleration region (that is, the region
occupied by ${\bf E} \parallel {\bf B}$ fields suitable for
curvature-dominated losses) is of order $R_S$; therefore both $R$ and $B$
are gouverned by a single parameter $M_{\rm BH}$, so one may express the
maximal energy through $M_{\rm BH}$ using results of
Sec.~\ref{sec:accel:regimes:eqns}.  Assuming
\begin{equation}
R\sim 5 R_S\approx 5 \times 10^{-5}~{\rm pc}~\frac{M_{\rm BH}}{10^8
M_\odot},
\label{Eq:R_S}
\end{equation}
one finds that for any reasonable $M_{\rm BH}$ (ranging from $\sim 10^6
M_\odot$ for normal galaxies through $(10^7 \dots 10^8) M_\odot$ for
Seyfert galaxies to $(10^9 \dots 10^{10}) M_\odot$ for powerful radio
galaxies and quasars) the maximal energy is determined by radiation losses
rather than by the Hillas condition and equals to
\begin{equation}
{\cal{E}}_{\rm max}={\cal{E}}_c\simeq 3.7 \times 10^{19}~{\rm eV}
\frac{A}{Z^{1/4}} \left(  \frac{M_{\rm BH}}{10^8 M_\odot}  \right)^{3/8}.
\label{Eq:Emax-MBH}
\end{equation}
This general constraint is presented in Fig.~\ref{fig:M-E_bh}
for different nuclei $(A,Z)$; for comparison, results of numerical
simulations of particle acceleration near a supermassive black
hole~\cite{Neronov2} are also plotted.

In a way similar to other observational manifestations of supermassive
black holes, both details of cosmic-ray acceleration and radiation losses
may depend on the accretion rate, accretion mode, environment etc.
However, we are interested here in the upper limit on the maximal
attainable energy of a cosmic-ray particle which is determined by $M_{\rm
BH}$ as we have just demonstrated.

\subsubsection{ Jets and outflows of active galaxies.}
\label{sec:accel:mf:jets}
Active galactic nuclei fuel large-scale (from sub-parsec to kiloparsec and
even Megaparsec length) extended more or less linear jets.
Revolutionary progress in the angular
resolution of radio (sub-milliarcsecond) and X-ray (sub-arcsecond) imaging
allowed for detailed studies and modelling of physical conditions in jets.
We briefly review basics of the current understanding of jet properties
following Refs.~\cite{0607228, 0604219}.

Seyfert galaxies possess extended structures which are often
non-collimated (opening angle of $45^\circ$ or more) and are found to be
non-relativistic; they are sometimes determined as ``outflows'', reserving
the term ``jets'' to strongly collimated relativistic flows. X-ray
emission from these outflows, when present, is well described by
thermal radiation (sometimes associated with star-forming regions in the
outflow \cite{0204361}).

\begin{figure}
\begin{center}
\includegraphics [width=0.7\textwidth]{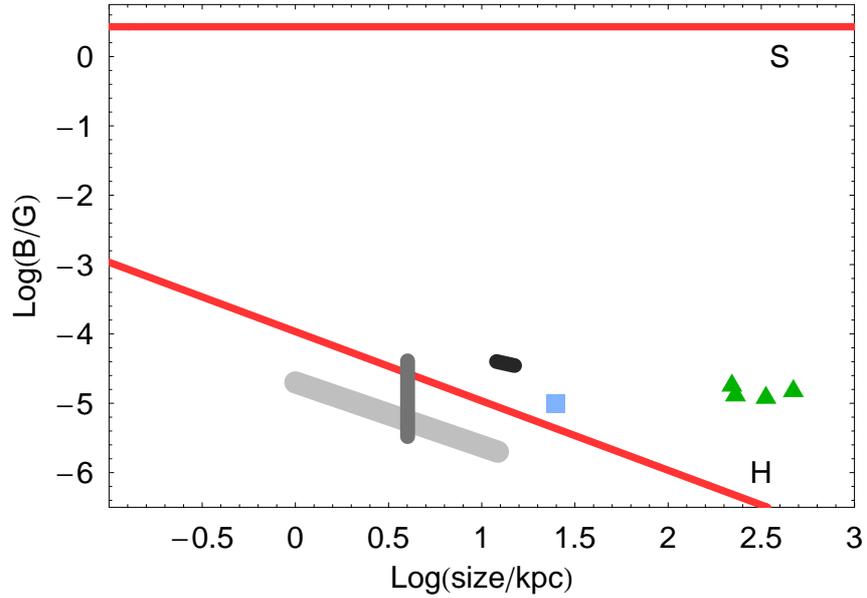}
\end{center}
\caption{
\label{fig:jets}
The size-field diagram for jets and outflows
of individual active galaxies.
Grey colors correspond to Seyfert galaxies, data from Refs.~\cite{0604219}
(light-grey diagonal line), \cite{ApJ-511-686} (grey vertical error bar),
\cite{0709.4476} (short dark grey diagonal). Blue box corresponds to FR~I
radio galaxy~\cite{ApJ-273-128}; green triangles represent quasar
jets~\cite{0306317}. The allowed region for acceleration of $10^{20}$~eV
protons is located between thick red lines H and S (the lower line, H,
corresponds to the Hillas limit; the upper one, S, corresponds to the
radiation-loss limit for inductive acceleration with synchrotron-dominated
losses). }
\end{figure}

Relativistic jets reveal themselves in non-thermal X-ray emission studied
now in great detail. The jets
are spatially resolved into components; in nearby jets (Cen~A) inner and
outer layers and bright knots are resolved. It is often assumed that
all jets are fuelled by the central black hole; the energy
flux is dominated by the magnetic-field energy at sub-parsec scales but
becomes particle-dominated at parsec scales. The emission of
low-luminosity sources (FR~I radio galaxies and BL~Lacs) is adequately
described by the synchrotron models from radio to X rays; their jets are
decelerated by entrainment of gas and dissipate in the end. High-power
FR~II and quasar jets bring their energy flux directly to their terminal
hot spots and require additional (e.g.\ Compton) component to describe
their spectra. Comparison of radio to X-ray observations gives rather firm
evidence to the origin of the emission of FR~I jets from accelerated
particles and to acceleration of these particles not only in a finite
number of shocks but also by means of some distributed mechanism along the
jet~\cite{0510661, 0710.1277}. Quite rarely, relativistic jets are present
in exceptionally powerful Seyfert galaxies; in these cases, they have
properties very similar to FR~I jets~\cite{0604219}. Models of
multifrequency spectra allow to constrain the magnetic field, the key
parameter of the synchrotron radiation.
The estimates depend also on the electron density; the degeneracy is often
removed either by the equipartition assumption or by a simultaneous
measurement of the self-Compton component, when applicable. When error
bars are given, they include the corresponding uncertainties. Some of these
estimates~\cite{0604219, ApJ-511-686, 0709.4476, ApJ-273-128, 0306317} are
presented in Fig.~\ref{fig:jets}.
In some cases, existence of ordered fields through the jet was proven, so
that the inductive acceleration may be possible (see e.g.\
Ref.~\cite{0106530}).

\subsubsection{Jet knots, hot spots and lobes of powerful active galaxies.}
\label{sec:accel:mf:hs-lobes}
\begin{figure}
\begin{center}
\includegraphics [width=0.7\textwidth]{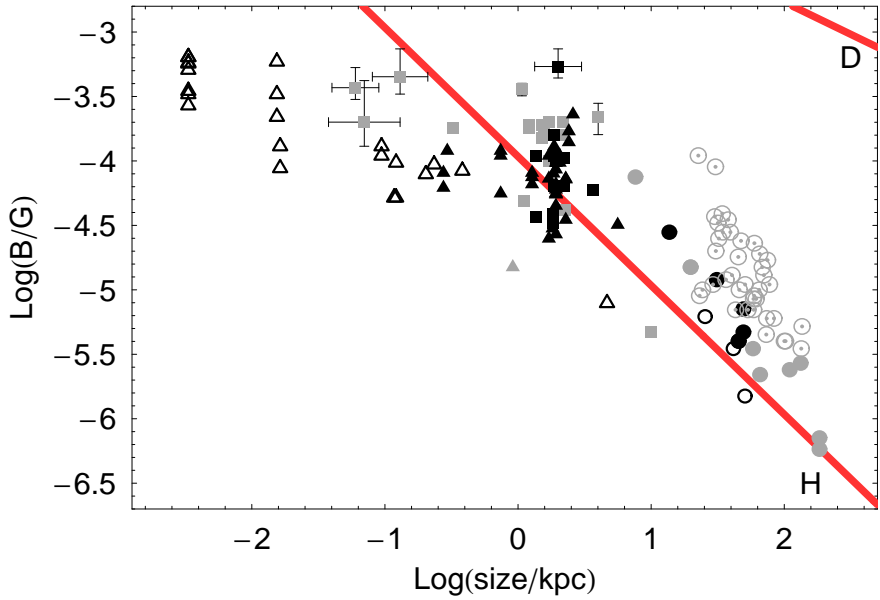}
\end{center}
\caption{
\label{fig:lobes}
The size-field diagram for knots (triangles), hot spots (boxes) and lobes
(circles) of individual powerful active galaxies. Filled black symbols
correspond to quasars and blazars, filled grey symbols correspond to FR~II
radio galaxies, empty symbols correspond to FR~I radio galaxies (data from
Ref.~\cite{KataokaStawarz}, X-ray observations assuming equipartition);
boxes with error bars represent the ``best-guess'' estimates of
Ref.~\cite{Meisenheimer}; dotted grey circles correspond to FR~II lobes
studied in Ref.~\cite{HardcastleFRII} (comparison of radio and X-ray
observations without the equipartition assumption).
The allowed region for acceleration of $10^{20}$~eV protons is located
between thick red lines H and D (the lower line, H, corresponds to the
Hillas limit; the upper one, D, corresponds to the radiation-loss limit for
diffusive acceleration). }
\end{figure}

When a relativistic jet is present, it may be accompanied by internal
shock regions (knots), terminal shock regions (hot spots) and extended
regions in the intergalactic space fuelled by the jet after its
termination (lobes). These regions are typically absent in low-power active
galaxies (Seyfert galaxies): knots are observed mostly in jets of FR~I
radio galaxies and quasars, lobes are typical for radio galaxies, hot
spots are present in the most powerful FR~II radio galaxies and quasars.
Magnetic fields may be determined either by X-ray synchrotron observations
alone (assuming equipartition) or by combined multifrequency observations
of both synchrotron and Compton radiations (allowing to relax the
equipartition assumption which occurs, in the end, a good approximation,
see e.g.\ Ref.~\cite{ApJ-612-729})\footnote{An interesting approach to
determination of the magnetic field in a knot in M87~\cite{0904.3925}
exploits the energy dependence of the energy loss rate, assuming it is
synchrotron-dominated. The resulting $\sim 0.6$~mG field is in a good
agreement with equipartition-based estimates.}. A summary of measurements
\cite{KataokaStawarz, Meisenheimer, HardcastleFRII} is given in
Fig.~\ref{fig:lobes}.

\subsection{ Star formation regions and starburst galaxies}
\label{sec:accel:mf:starburst}
\begin{figure}
\begin{center}
\includegraphics [width=0.7\textwidth]{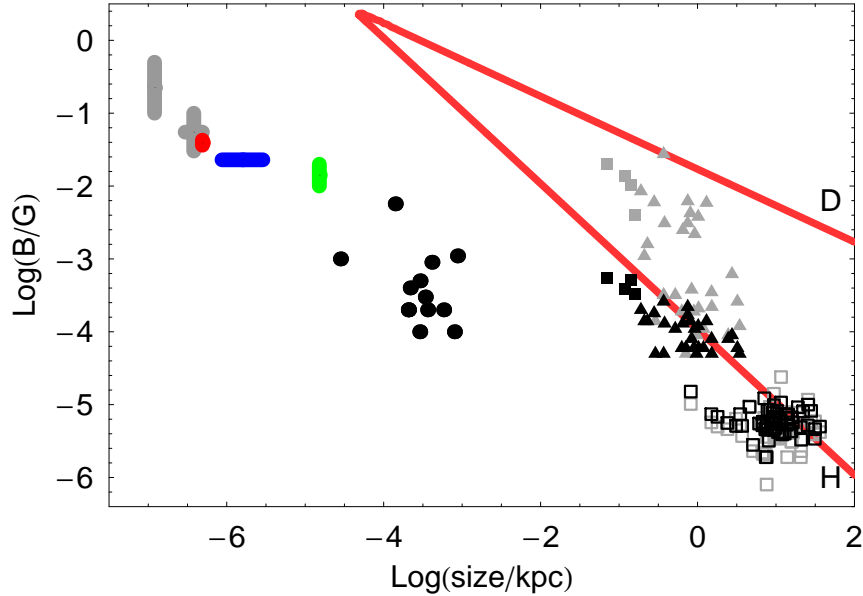}
\end{center}
\caption{
\label{fig:starburst}
The size-field diagram for Galactic star-forming regions ($R\lesssim$~pc)
and starburst galaxies ($R \gtrsim 0.1$~kpc). Thick (error-bar) lines
correspond to measurements of the Zeeman splitting in masers (grey,
Ref.~\cite{0510452}; red, Ref.~\cite{0605741}; blue,
Ref.~\cite{0804.1141}; green, Ref.~\cite{ApJ-674-295}). Black dots
represent results of submillimeter imaging polarimetry of
Ref.~\cite{0709.0256}. Data for normal (empty boxes), starburst
(triangles) and extreme starburst (filled boxes) galaxies are taken from
Ref.~\cite{WaxmanSB}; black symbols correspond to the minimal-energy field
estimates while grey symbols correspond to equipartition field estimates.
The allowed region for acceleration of $10^{20}$~eV protons is located
between thick red lines H and D (the lower line, H, corresponds to the
Hillas limit; the upper one, D, corresponds to the radiation-loss limit for
diffusive acceleration). }
\end{figure}

Measurements of the magnetic field in Galactic star-forming regions
becomes possible with the Zeeman splitting in masers in circumstellar
disks~\cite{0510452, 0605741, 0804.1141, ApJ-674-295} and infrared imaging
polarimetry~\cite{0709.0256}. Though these regions in our Galaxy have
never been considered as possible sites of UHECR acceleration, these
measurements may give some hints to the fields in larger star-forming
regions in starburst galaxies, where particles could be accelerated to
very high energies e.g.\ in shocks from subsequent supernova explosions
\cite{9903145}; magnetic fields in these extragalactic sites are measured
indirectly. A summary of measurements is given in
Fig.~\ref{fig:starburst};
a number of arguments in favour of higher (equipartition) fields in
starburst galaxies were presented in Ref.~\cite{WaxmanSB} while continuity
(see Fig.~\ref{fig:starburst}) with
the Galactic measurements may support lower (minimal-energy)
estimates.

\subsection{ Gamma-ray bursts}
\label{sec:accel:mf:grb}
Estimates of
the magnetic field in gamma-ray
bursts (GRBs) assume~\cite{GRB:Piran} that the origin of both prompt and
afterglow emissions in a certain part of the spectrum is the synchrotron
radiation of relativistic electrons.  This assumption is supported by
measurements of the afterglow spectra and lightcurves and by observation
of the strongly polarized prompt emission (see Ref.~\cite{GRB:Piran} for
discussion and references). Ref.~\cite{GRB:Piran} quotes $B \sim 10^6$~G
for $R\sim (10^{13} \div 10^{15})$~cm (prompt emission) and $B \sim 1$~G
for $R\sim (10^{16} \div 10^{18})$~cm (afterglow) (we assume that the
estimates correspond to the observer's rest frame). Another, somewhat higher
field estimate may be obtained following Ref.~\cite{Derishev} (see also
Ref.~\cite{0409489}) from the total luminosity of a GRB, assuming that the
magnetic-field energy ${\cal{E}}_m$ is a fraction $\epsilon _m < 1$ of the
radiation energy ${\cal{E}}_{\rm rad}$. However, this estimate depends
strongly on the assumed beaming.

\begin{figure}
\begin{center}
\includegraphics [width=0.7\textwidth]{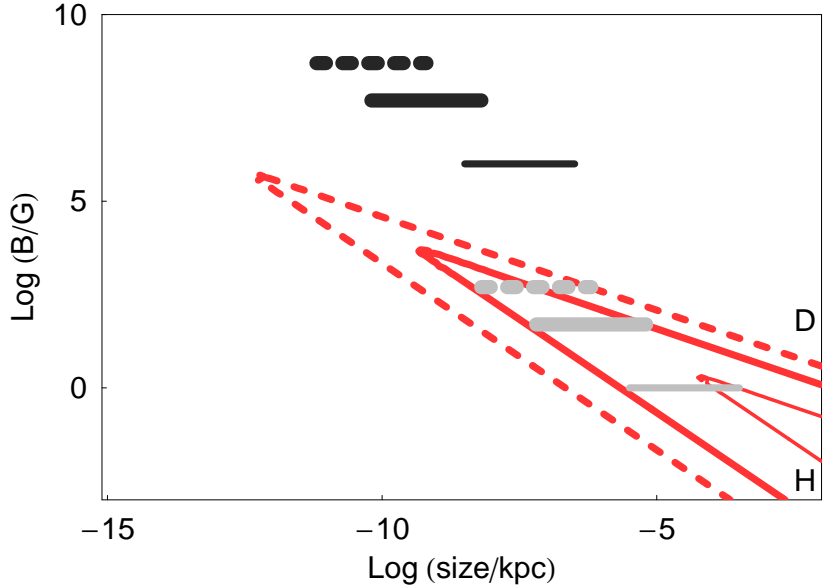}
\end{center}
\caption{
\label{fig:grb}
The size-field diagram for gamma-ray bursts.
Horizonthal lines represent estimates of Ref.~\cite{GRB:Piran}
which assume the synchrotron origin for the prompt emission (dark grey)
and the afterglow (light grey). The allowed region for acceleration of
$10^{20}$~eV protons is located between
(the lower lines, H, correspond
to the Hillas limit; the upper ones, D, correspond to the radiation-loss
limit for diffusive acceleration).
Dashed lines assume $\Gamma=500$, thick
lines assume $\Gamma=50$,
thin
lines assume $\Gamma=1$.
}
\end{figure}

Within the scope of this paper, we may estimate the maximal energy
${\cal{E}}_{\rm max}$ of accelerated particles in the comoving frame
following equations of Sec.~\ref{sec:accel:regimes:eqns}
for shock (diffusive) acceleration. The GRB shells are however
ultrarelativistic ($\Gamma \sim 100$, see e.g.\ Ref.~\cite{GRBGamma}) and
we have to multiply the comoving-frame ${\cal{E}}_{\rm max}$ by $\Gamma$
to get the maximal rest-frame energy. Results are presented in
Fig.~\ref{fig:grb}
which, for the GRB case, is more instructive than the summary plots of
Sec.~\ref{sec:accel:summary}.  We note that at large $\Gamma$,
the maximal energy may be limited by interactions with thermal
photon field (not taken into account in the present work) and {\em
decreases} as $\Gamma^{-1}$ at large $\Gamma$~\cite{Derishev}.

\subsection{ Galaxy clusters, superclusters and voids}
\label{sec:accel:mf:clusters}
Information about cluster magnetic fields comes mostly from observations
of their extended radio, and sometimes X-ray, emission. These observations
are reviewed e.g.\ in Refs.~\cite{Massimo, 0410182, 0801.0985}, where more
references to original works may be found. Estimates based on
equipartition (see e.g.\ Refs.~\cite{0507367, 0704.3288}), as well as
those
assuming Compton scattering on CMB photons, favour values of $B\sim (0.1
\div 1)~\mu$G at megaparsec scales; Faraday rotation measurements (see
e.g.\ Refs.~\cite{0011281, 0602622, 0608433}) favour somewhat higher
fields,
$B\sim (1 \div 5)~\mu$G. Model-dependent numerical simulations remain the
main source of information about magnetic fields at the supercluster
scales ($R\sim 100$~Mpc), escepially in voids. Estimates vary between
$B\sim 10^{-11}$~G \cite{DolagMF} and $B\sim 10^{-8}$~G \cite{SiglMF}.

\section{Summary and discussion}
\label{sec:accel:summary}
\begin{figure}
\begin{center}
\includegraphics [width=0.7\textwidth]{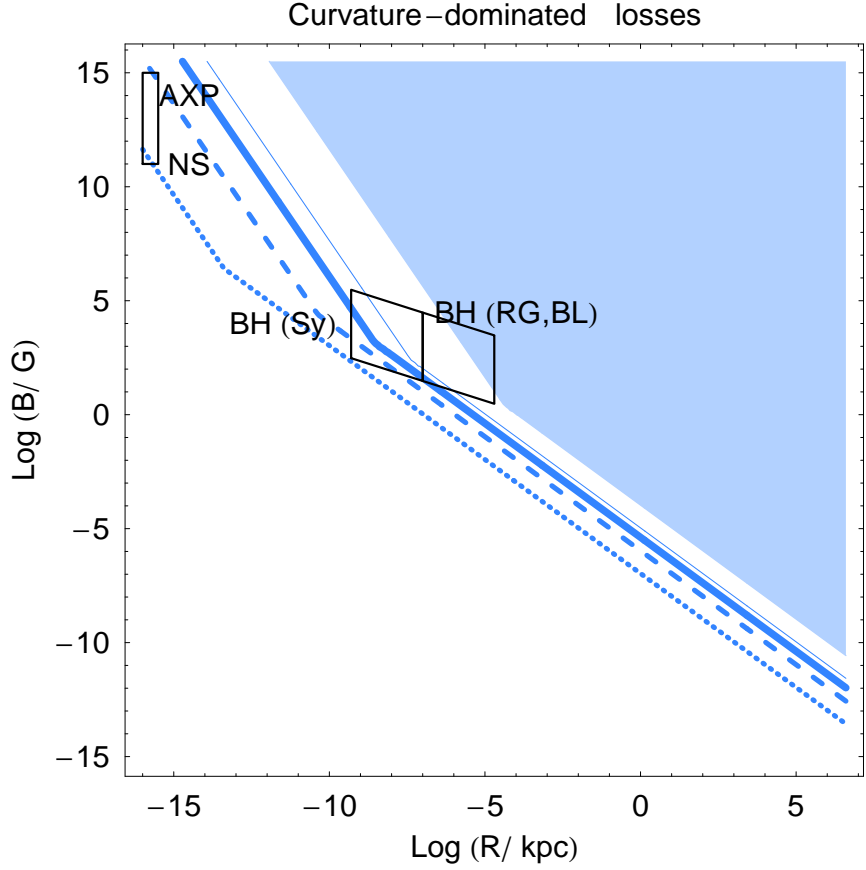}
\end{center}
\caption{
\label{fig:HillasCurv}
The size-field plot with constraints from geometry and radiation losses
for the regime where losses are dominated by curvature radiation. These
are minimal possible losses and these constraints are therefore the most
weak. Boxes denote parameter regions for objects in which conditions for
this loss regime may be satisfied, that is immediate neighbourhood of
neutron stars (NS), anomalous X-ray pulsars and magnetars (AXP) and of
supermassive central black holes (BH) of active galactic nuclei, from
low-power Seyfert galaxies (Sy) to powerful radio galaxies (RG) and
blazars (BL). The shaded area corresponds to the parameter region where
acceleration of protons to $10^{20}$~eV is possible. Lines bind from
below the allowed regions for $10^{19}$~eV protons (thin full line),
$10^{20}$~eV iron nuclei (thick full line), $10^{18}$~eV protons (dashed
line) and $10^{17}$~eV protons (dotted line). Right-hand parts of the
lines represent the Hillas constraint while left-hand (steeper) parts
represent the radiation-loss constraint.}
\end{figure}

\begin{figure}
\begin{center}
\includegraphics [width=0.7\textwidth]{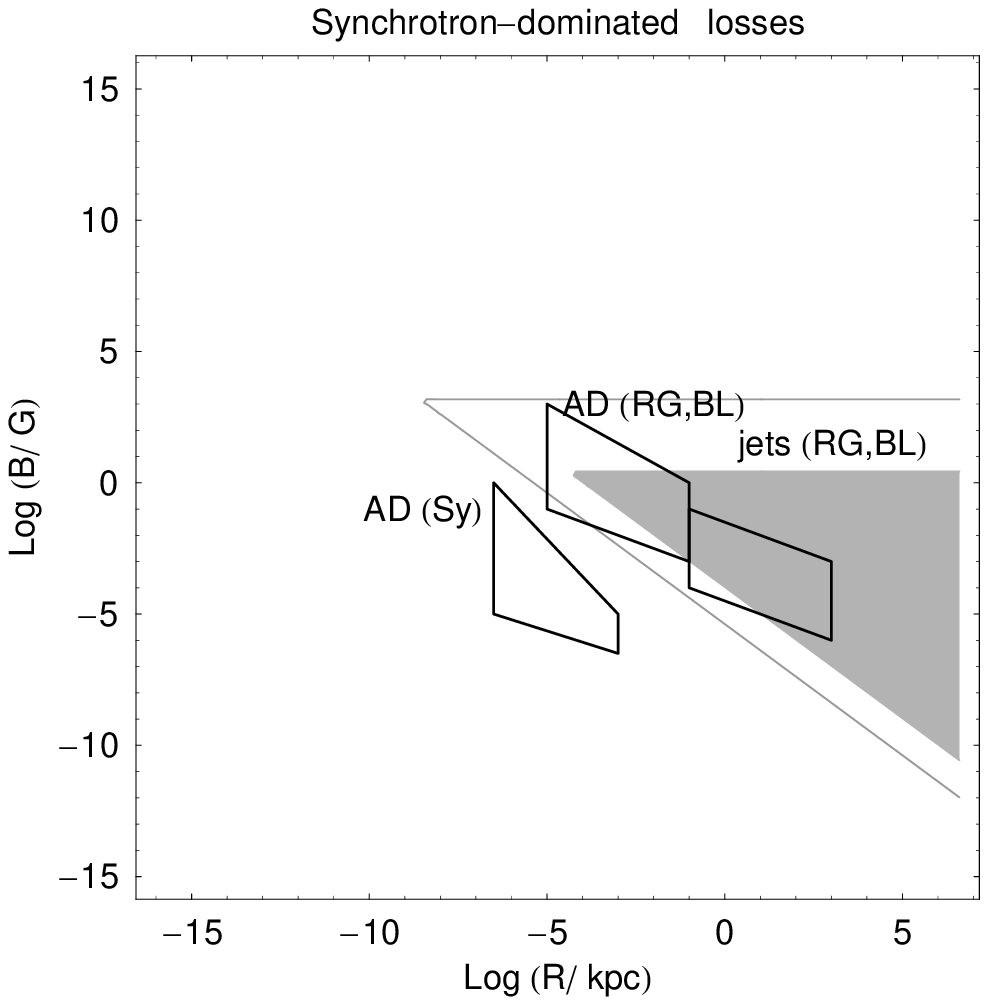}
\end{center}
\caption{
\label{fig:HillasSync}
The size-field plot with constraints from geometry and radiation losses
for the regime of one-shot acceleration with synchrotron-dominated losses.
Boxes denote parameter regions for objects in which conditions for this
loss regime may be satisfied, that is central parsecs (AD) of active
galaxies (low-power Seyfert galaxies (Sy) and powerful radio galaxies
(RG) and blazars (BL)) and relativistic jets of powerful active galaxies.
The shaded area corresponds to the parameter
region where acceleration of protons to $10^{20}$~eV is possible.
The line binds the allowed regions for $10^{20}$~eV iron nuclei.
Lower lines represent the Hillas
constraint while upper (horizonthal) lines represent the radiation-loss
constraint.
All quantities are given in the comoving frame, so the maximal energy
for jets should be multiplied by the bulk Lorentz factor of the jet
which may be as large as
$\sim 10$ for leptonic jets and $\sim 100$ for hadronic
jets~\cite{Derishev}.
}
\end{figure}

\begin{figure}
\begin{center}
\includegraphics [width=0.7\textwidth]{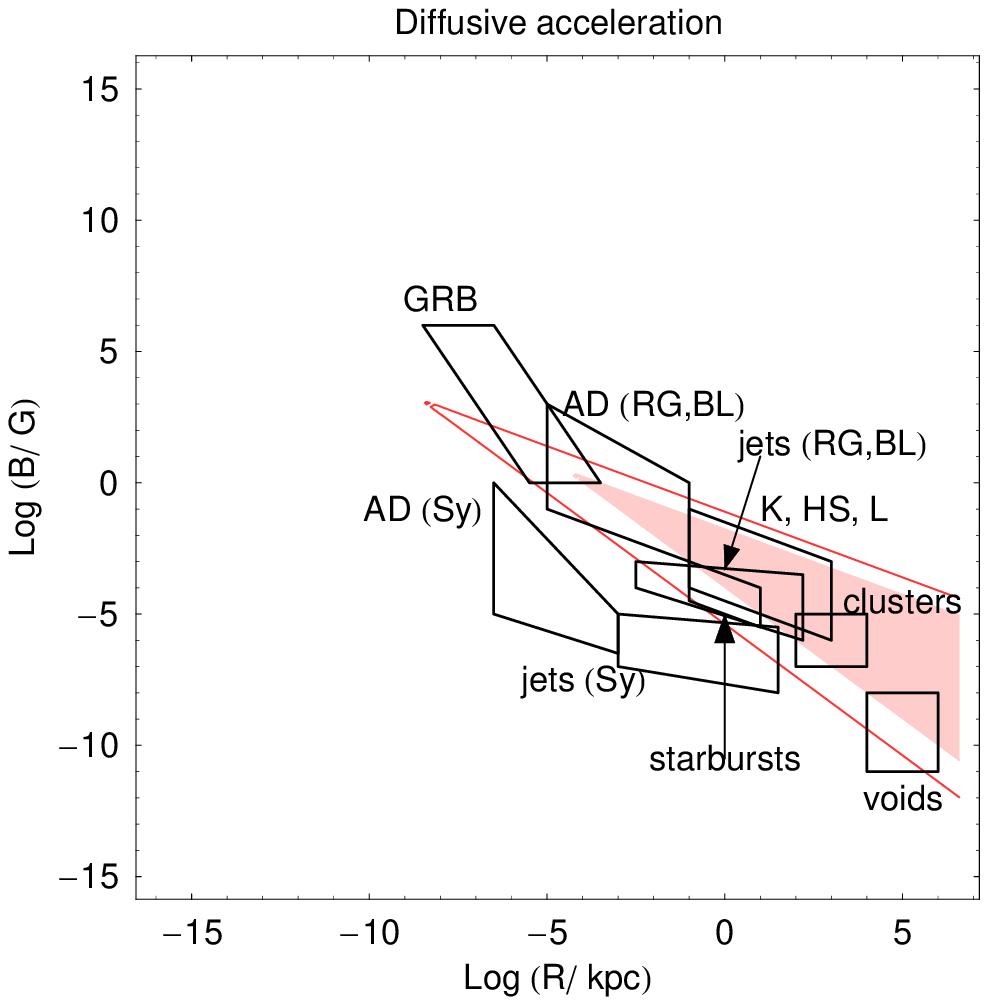}
\end{center}
\caption{
\label{fig:HillasDiffuse}
The size-field plot with constraints from geometry and radiation losses
for the regime of diffusive acceleration with synchrotron-dominated losses.
Boxes denote parameter regions for objects in which conditions for this
loss regime may be satisfied, that is central parsecs (AD) of active
galaxies (low-power Seyfert galaxies (Sy) and powerful radio galaxies (RG)
and blazars (BL)), relativistic jets, knots (K), hot spots (HS) and
lobes (L) of powerful active galaxies (RG and BL); non-relativistic jets of
low-power galaxies (Sy); starburst galaxies; gamma-ray bursts (GRB);
galaxy clusters and intercluster voids. The shaded area corresponds to the
parameter region where acceleration of protons to $10^{20}$~eV is
possible. The line binds the allowed regions for $10^{20}$~eV iron nuclei.
Lower lines represent the Hillas constraint while upper
lines represent the radiation-loss constraint. All quantities are given in
the comoving frame, so the maximal energy for jets and shells of GRBs
should be multiplied by the bulk Lorentz factor which may be as large as
$\sim 10$ for leptonic jets and $\sim 100$ for hadronic jets and
GRBs~\cite{Derishev}. }
\end{figure}

\begin{figure}
\begin{center}
\includegraphics [width=0.7\textwidth]{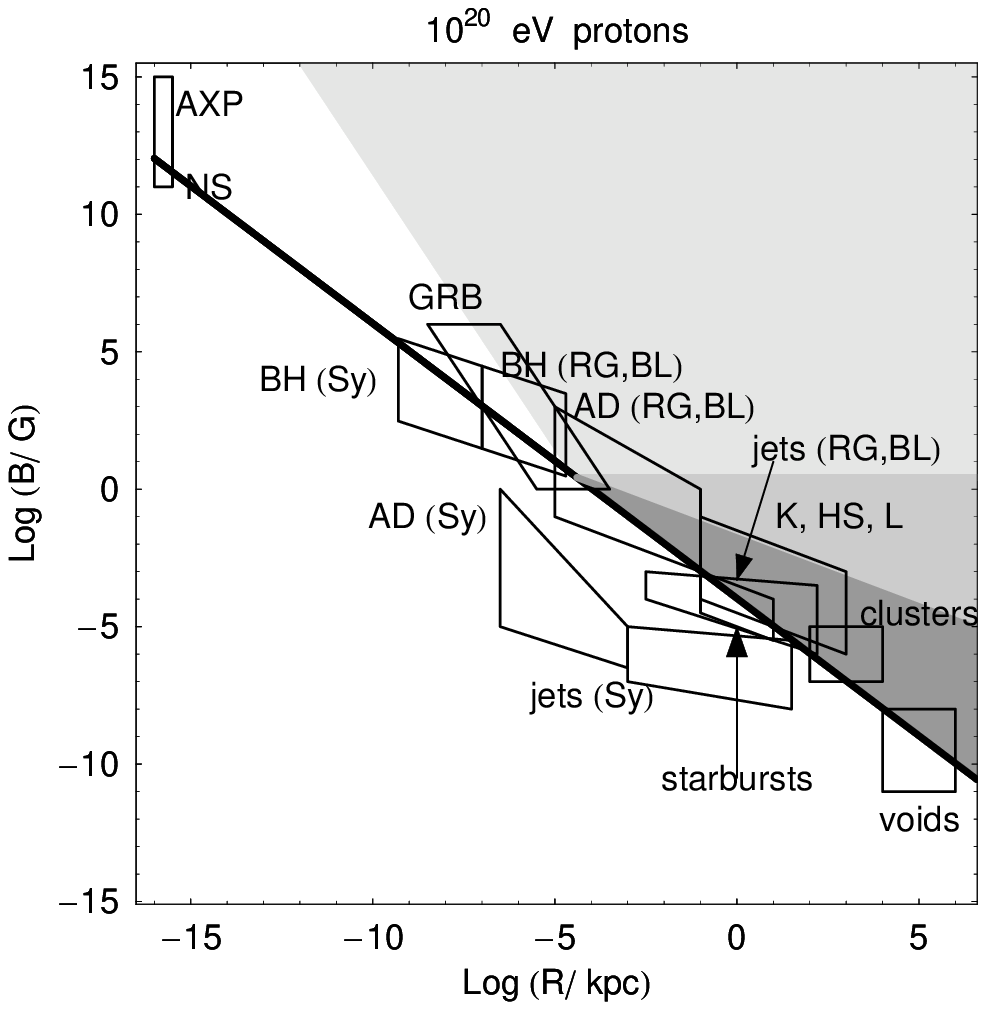}
\end{center}
\caption{
\label{fig:HillasP}
The Hillas plot with constraints from geometry and radiation losses for
$10^{20}$~eV protons. The thick line represents the lower boundary of the
area allowed by the Hillas criterion. Shaded areas are allowed by the
radiation-loss constraints as well: light grey corresponds to one-shot
acceleration in curvature-dominated regime only; grey allows also for
one-shot acceleration in synchrotron-dominated regime; dark grey allows
for both one-shot and diffusive (e.g.\ shock) acceleration. See captions
to Figs.~\ref{fig:HillasCurv}, \ref{fig:HillasSync},
\ref{fig:HillasDiffuse} for notation of boxes corresponding to potential
sources.
}
\end{figure}

\begin{figure}
\begin{center}
\includegraphics [width=0.7\textwidth]{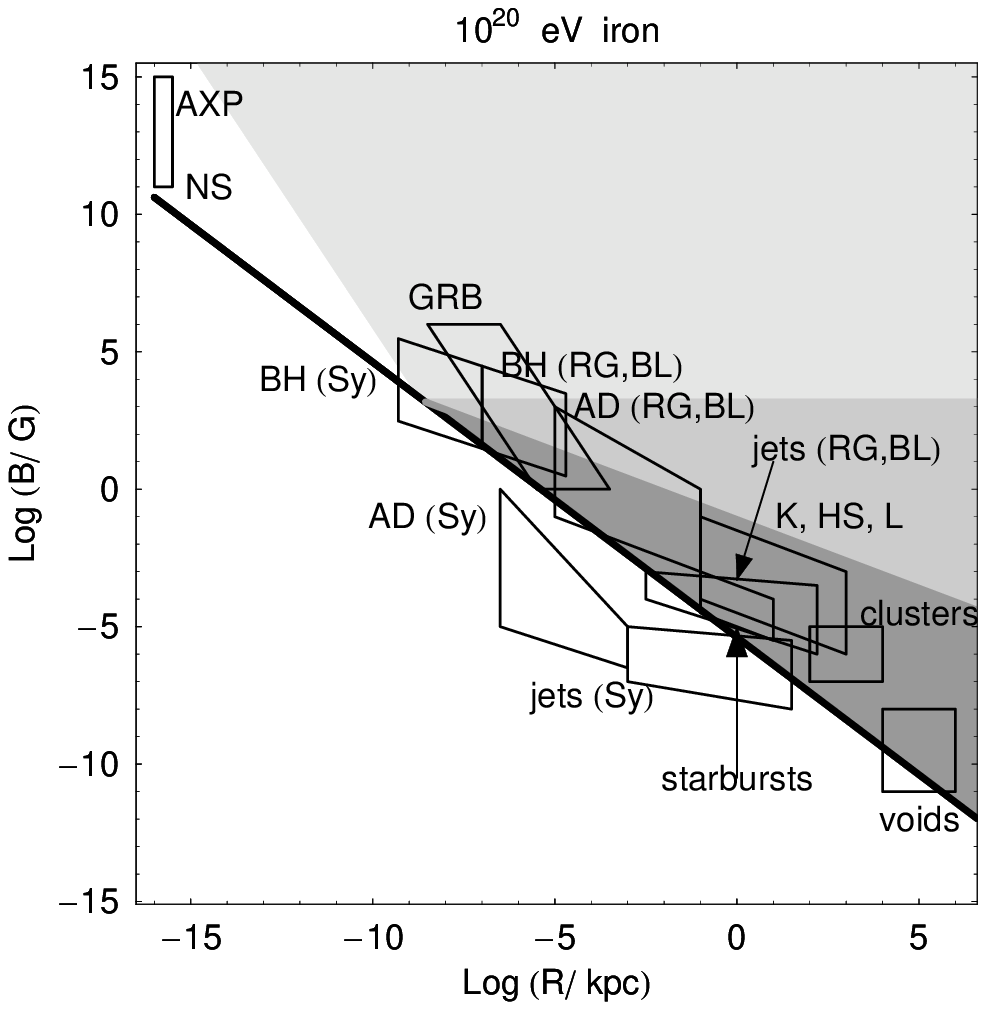}
\end{center}
\caption{
\label{fig:HillasFe}
The same as Fig.~\ref{fig:HillasP} but for $10^{20}$~eV iron nuclei. The
most important difference with Fig.~\ref{fig:HillasP} is that acceleration
of iron nuclei to $10^{20}$~eV is possible (unlike for protons) in
low-power active galaxies (e.g.\ Seyfert galaxies).}
\end{figure}

Based on the data collected in Sec.~\ref{sec:accel:mf-measurements} and on
the limits on the maximal energy, Sec.~\ref{sec:accel:regimes:eqns}, we
redraw here the Hillas plot supplemented by radiation-loss constraints.
Figures~\ref{fig:HillasCurv} -- \ref{fig:HillasDiffuse} give constraints
for particular acceleration regimes while Figs.~\ref{fig:HillasP},
\ref{fig:HillasFe} represent our updated summary Hillas plots.

The weakest possible constraints (for inductive acceleration with
curvature-dominated losses) are presented in Fig.~\ref{fig:HillasCurv}.
Constraints for inductive acceleration with synchrotron-dominated losses,
applicable mostly to inner and outer jets of active galaxies, are given in
Fig.~\ref{fig:HillasSync},
while constraints for the most general diffusive acceleration are
presented in Fig.~\ref{fig:HillasDiffuse}.
Figure~\ref{fig:HillasP} represents our version of the Hillas plot with
constraints for $10^{20}$~eV protons.
Figure~\ref{fig:HillasFe} is the
same plot but for $10^{20}$~eV iron nuclei.

Constraints for neutron stars follow from Sec.~\ref{sec:accel:mf:psr};
even for the least restrictive acceleration regime they are not satisfied
for UHE particles. In active galaxies, various regimes of acceleration may
operate. In the immediate vicinity of the central black hole (up to a few
$R_S$), the field configuration allows for the inductive acceleration with
curvature-dominated losses. These regions are denoted as ``BH'' in
Figs.~\ref{fig:HillasCurv}, \ref{fig:HillasP}, \ref{fig:HillasFe}; the
parameters correspond to Eqns.~\ref{Eq:R_S} and \ref{ZnajekLimit}. The
latter one is an upper limit on the field, so we extend the boxes for two
orders of magnitude lower in $B$, cf.\ Fig.~\ref{fig:M-B_bh}. Beyond a few
$R_S$, the ${\bf E} \parallel {\bf B}$ field structure no longer holds,
but coherent fields may still be present in inner jets. For these central
parsecs of AGN (denoted as ``AD'' in
Figs.~\ref{fig:HillasSync}--\ref{fig:HillasFe}) we use the field estimates
from Fig.~\ref{fig:bh-measurements}. For the extended parts of active
galaxies (jets, jet knots, hot spots and lobes) we use field estimates
summarized in Figs.~\ref{fig:jets}, \ref{fig:lobes}. The summary boxes for
starburst galaxies include both equipartition and minimal-energy estimates
(Fig.~\ref{fig:starburst}). For GRB, the summary plots present
synchrotron-based estimates for both inner and outer shocks (see
Fig.~\ref{fig:grb} for a more instructive plot). Field estimates for
clusters, superclusters and voids follow Sec.~\ref{sec:accel:mf:clusters}.

The constraints discussed here and expressed in terms of the Hillas plot
are necessary, but they should be supplemented by other limits (listed in
the beginning of Sec.~\ref{sec:accel:general}). We note that in estimation
of the maximal atteinable energy, important constraints are put by
interactions of accelerated particles with ambient photons. In particular,
interaction with the cosmic microwave background is important for large,
$R \gtrsim $Mpc, sources (lobes of radio galaxies, clusters and voids),
while interaction with the internal source radiation field is important
for ultraluminuous sources (GRB and AGN). These constraints, considered
elsewhere, further restrict the number of potential UHECR
accelerators\footnote{In certain cases the proton-gamma interactions,
instead of pure dissipation, can significantly amplify the acceleration
process \cite{DerishevStern}.}. For the diffusive shock acceleration,
these constraints have been studied e.g.\ in Ref.~\cite{Protheroe}.

The maximal energy for the supermassive black holes is readily expressed
in terms of a single parameter, the black-hole mass $M_{\rm BH}$. We used
the upper limit on the magnetic field, $B_{\rm BH}$, which is most likely
one or two orders of magnitude higher than the actual values, so that our
estimate, Eq.~(\ref{Eq:Emax-MBH}), is robust. It depends weakly
($\sqrt{R/R_S}$) on the assumed size of the acceleration region.

While we tried to make all constraints as robust as possible, it is clear
that they should be considered as order-of-magnitude estimates (in fact,
typical precision of the magnetic-field determination is an order of
magnitude) and, for individual unusual field configurations, can be
quantitatively violated. An example of such a configuration is a linear
accelerator with curvature radius $r$ of field lines exceeding the size
of the source $R$; then $R$ should be substituted by $r$ in
Eq.~(\ref{Eq:loss:curv}). The estimates should be used with care for the
cases when the magnetic field changes violently within the accelerator
(for instance when particles are accelerated by magnetic reconnection). A
more detailed modelling of acceleration and losses is required in these
cases.

One of our most important conclusions is that low-power active galaxies
(e.g.\ Seyfert galaxies) cannot accelerate protons to energies $\gtrsim
5\times 10^{19}$~eV. Indeed, when the extended structures (jets and
outflows) are present, the magnetic field there is far too weak to satisfy
the Hillas condition (even for very rare relativistic jets), see
Fig.~\ref{fig:jets}. The same is true for the accretion disks, where the
field is nicely constrained from above by non-observation of the Zeeman
splitting in megamasers (Fig.~\ref{fig:bh-measurements}). The most
favourable conditions for acceleration correspond to the immediate
vicinity (a few $R_S$) of the central black hole where the upper limit on
the maximal energy is given by Eq.~(\ref{Eq:Emax-MBH}). Since for Seyfert
galaxies, $M_{\rm BH}\lesssim (10^7 \div 10^8) M_\odot$, proton
acceleration to $\sim 10^{20}$~eV is not allowed. However, these (and only
these) central parts of Seyfert galaxies can, in principle, accelerate
protons to $\sim 10^{18}$~eV and heavy nuclei to $\sim 10^{20}$~eV, if
interactions with ambient photons are weak enough. Though heavy nuclei are
much less abundant than protons, Seyfert galaxies themselves are much more
abundant, and hence typically close to the observer, than powerful radio
galaxies and blazars, so that their population can contribute to the UHECR
spectrum.

\section{Conclusions}
\label{sec:concl}

We reviewed constraints on astrophysical UHE accelerators and presented
the Hillas plot supplemented with radiation-loss constraints and updated
with recent astrophysical data. Contrary to previous studies, we emphasised
that active galaxies span a large area on the plot, and only the most
powerful ones (radio galaxies, quasars and BL Lac type objects) are
capable of acceleration of protons to UHE. If UHECR particles are
accelerated close to the supermassive black holes in AGN, then most likely
the mechanism is ``one-shot'' with energy losses dominated by the
curvature radiation.
Other potential UHE
acceleration sites are jets, lobes, knots and hot spots of {\em powerful}
active galaxies, starburst galaxies and shocks in galaxy clusters.
Acceleration of particles in supercluster-scale shocks, gamma-ray bursts
and inner part of AGN is subject to additional constraints from $p \gamma$
interactions which are not discussed here.

Unlike protons, heavy nuclei can be accelerated to UHE in circumnuclear
regions of low-power active galaxies. Since these galaxies are abundant,
this contribution to UHECR flux may be important, leading to a mixed
primary cosmic-ray composition at highest energies.

The authors are indebted to D.~Gorbunov, S.~Gureev, A.M.~Hillas, V.~Lukash,
A.~Neronov, S.~Popov and D.~Semikoz for interesting discussions. We
acknowledge the use of online tools~\cite{LEDA, NED}. This work was
supported in part by the grants
RFBR 07-02-00820 09-07-08388 (ST), NS-1616.2008.2 (ST) and by FASI
government contracts 02.740.11.0244 (ST) and
02.740.11.5092 (KP, ST) and by the Dynasty Foundation (KP, ST).
.

\appendix

\section{Derivation of electrodynamic results}
\label{app:ed}
\subsection{Energy losses for the curvature radiation (see e.g.\
Ref.~\cite{Ginz})}
\label{app:curv}
Consider a particle moving along
curved field lines. The particle has a longitudinal velocity
component (${\bf v}_\parallel \parallel {\bf B}$) and a drift
component (${\bf v}_{\rm d}\perp{\bf B}$). This drift component
provokes the appearance of the Lorentz force which curves the
particle's trajectory towards the field lines. For a relativistic
particle,
$$v_{\rm d}=\frac{{v_{\parallel}}^{2}m}{qBr}
{\left(\frac{\cal{E}}{m}\right)},
$$
so the Lorentz force is
$${\bf F}_{\rm L}=q\bigl[{\bf v}_{\rm d}\times{\bf B}\bigr],$$
$$F_{\rm L}=
\frac{v_\parallel^2 m}{r}
\left(
\frac{\cal{E}}{m}
\right).
$$
The energy losses are in general determined by Eq.~(\ref{5*}) which may be
rewritten as
$$
\frac{d{\cal{E}}}{dt}=
\frac{2q^{2}}{3m^{2}\left(1-{v^{2}}\right)}\left[F^{2}-{\left({\bf
F}\cdot {\bf v}\right)}^{2}\right].
$$
In the regime we consider,
$\left({\bf F}\cdot {\bf v}\right)=0$ and
consequently
$$
\frac{d{\cal{E}}}{dt}=
\frac{2q^{2}{v_{\parallel}}^{4}}{3r^{2}}
{\left(\frac{\cal{E}}{m}\right)}^{4}.$$
In the ultrarelativistic limit $v_{\parallel}\rightarrow c$ one obtains
Eq.~(\ref{Eq:loss:curv}).

\subsection{The maximal energy for diffusive acceleration~\cite{Medvedev}}
\label{app:medv}

Consider a flow propagating through a
magnetized medium. An accelerated particle gains energy by
repeated scattering off the flow. After every
scattering, the particle travels along the Larmor
orbit,
radiates and slows
down according to Eq.(\ref{Eq:loss:sync}); consequently
$$
\int^{{\cal{E}}}_{{\cal{E}}_{0}}
\frac{d{\cal{E}}}{{\cal{E}}^{2}}=
-\frac{2q^{4}}{3m^{4}}\int^{R}_{0}B^{2}(x)dx ,
$$
hence
$$
\frac{1}{{\cal{E}}}=
\frac{1}{{\cal{E}}_{0}}+
\frac{2q^{4}}{3m^{4}}\int^{R}_{0}B^{2}(x)dx.
$$
The maximal energy ${\cal{E}}={\cal{E}}_{\rm cr}$ is determined by setting
${\cal{E}}_0 \to \infty$,
$$\frac{1}{{\cal{E}}_{\rm cr}}=
\frac{2q^{4}}{3m^{4}}\int^{R}_{0}B^{2}(x)dx
\simeq
\frac{2q^{4}}{3m^{4}}B^2R,
$$
and we obtain Eq.~(\ref{6*}).


\begin{thebibliography}{89}

\bibitem{NaganoWatson}
Nagano M and Watson A A,
{\it Observations And Implications Of The Ultrahigh-Energy Cosmic Rays,}
2000, {\it  Rev.\ Mod.\ Phys.}\  {\bf 72} 689

\bibitem{Kachelriess:lectures}
Kachelriess M,
2008,
  arXiv:0801.4376 [astro-ph]

\bibitem{G}
Greisen K,
1966
{\it    Phys.\ Rev.\ Lett.}\
  {\bf 16} 748

\bibitem{ZK}
Zatsepin G T and Kuzmin V A,
1966
{\it    JETP Lett.}\
  {\bf 4} 78

\bibitem{HiRes:cutoff}
Abbasi R {\it et al.},
  (The High Resolution Fly's Eye  Collaboration),
2008,
  {\it Phys.\ Rev.\ Lett.}\ {\bf 100} 101101

\bibitem{PAOspectrum}
The Pierre Auger Collaboration,
2008,
  arXiv:0806.4302 [astro-ph]

\bibitem{StellarMagnetospheres}
Schl\"uter A und Biermann L,
1950, {\it Z.\ Naturforsch.}\ {\bf 5a}
237

\bibitem{Hillas}
Hillas A M,
1984,
{\it Ann.\ Rev.\ Astron.\ Astrophys.}\ {\bf 22} 425

\bibitem{Protheroe}
Protheroe R J,
2004,
  {\it Astropart.\ Phys.}\  {\bf 21} 415

\bibitem{Derishev}
Aharonian F A,
2002,
  {\it Phys.\ Rev.}\  {\bf D66} 023005

\bibitem{Medvedev}
Medvedev M V,
2003,
 {\it Phys.\ Rev.}\   {\bf E67} 045401

\bibitem{paper2}
Gureev S and Troitsky S,
arXiv:0808.0481 [astro-ph]

\bibitem{Semikoz:spectrum}
Kachelriess M and Semikoz D V,
2006 {\it Phys.\ Lett.}\   {\bf B634} 143

\bibitem{LL}
Landau L and Lifshitz E,
{\it The classical theory of fields}, 1951, Addison--Wesley

\bibitem{Longair}
Longair M S,
{\it High-energy astrophysics. Vol. 1: Particles, photons and their
detection,}
1992, Cambridge Univ. Press

\bibitem{Fermi}
Fermi E,
1949
{\it Phys.\ Rev.}\  {\bf 75} 1169

\bibitem{BlandEich}
Blandford R and Eichler D,
1987
{\it  Phys.\ Rept.}\  {\bf 154} 1

\bibitem{OstrLayers}
Rieger F M and Duffy P,
2004 {\it Astrophys.\ J.}\  {\bf 617}  155

\bibitem{DerishevStern}
Derishev E V, Aharonian F A, Kocharovsky V V and Kocharovsky Vl V,
2003
{\it Phys.\ Rev.}\  D {\bf 68} 043003

\bibitem{Ostrowski08}
Ostrowski M,
  2008, arXiv:0801.1339 [astro-ph]

\bibitem{0106530}
Schopper R, Birk G T and Lesch H,
2001 {\it Astropart.\ Phys.}\  {\bf 17} 347

\bibitem{PSR}
Venkatesan A, Miller M C and Olinto A V,
1997
  {\it Astrophys.\ J.}\  {\bf 484}  323

\bibitem{Neronov0}
Neronov A and Semikoz D,
2003
{\it New Astron.\ Rev.}\ {\bf 47} 693

\bibitem{Neronov1}
Neronov A, Tinyakov P and Tkachev I,
2005
  {\it J.\ Exp.\ Theor.\ Phys.}\  {\bf 100} 656

\bibitem{Neronov2}
  Neronov A, Semikoz D and Tkachev I,
2007, arXiv:0712.1737 [astro-ph]

\bibitem{sources}
Torres D F and Anchordoqui L A,
2004
{\it    Rept.\ Prog.\ Phys.}\
  {\bf 67} 1663
  [arXiv:astro-ph/0402371]

\bibitem{comparative}
Gorbunov D and Troitsky S,
2005 {\it    Astrop.\ Phys.}\ {\bf 23} 175

\bibitem{Massimo}
Giovannini M,
2004
  {\it Int.\ J.\ Mod.\ Phys.}\  D {\bf 13} 391

\bibitem{Vallee}
Valle\'e J P,
2004,
{\it New Astron.\ Rev.}\ {\bf 48} 763

\bibitem{0804.0250}
Mereghetti S,
  arXiv:0804.0250 [astro-ph]

\bibitem{Nature-423-725}
  Bignami G F
  {\it et al.},
2003 {\it Nature} {\bf 423} 725

\bibitem{0610382}
Baring M G and Harding A K,
  2007 {\it Astrophys.\ Space Sci.}\  {\bf 308}  109

\bibitem{Carroll}
Carroll B W and Ostlie D A,
{\it An introduction to modern astrophysics},
2007, Pearson/Addison Wesley

\bibitem{Postnov}
Zasov V A and Postnov K A,
{\it General astrophysics},
2006, Vek-2 (in Russian)

\bibitem{Veron-classif}
V\'eron-Cetty M P and V\'eron P,
2000
{\it Astron.\ Astrophys.\ Rev.}\ {\bf 10} 81

\bibitem{FR}
Fanaroff B L and Riley J M,
1974
{\it Mon.\ Not.\ Roy.\ Astron.\ Soc.}\
  {\bf 167} 31P

\bibitem{0610912}
Vlemmings W H T, Bignall H E and Diamond P J,
2007
  {\it Astrophys.\ J.}\  {\bf 656} 198

\bibitem{0502240}
Modjaz M
  {\it et al.},
  J.~M.~Moran J M, Kondratko P T and Greenhill L J,
2005
  {\it Astrophys.\ J.}\  {\bf 626} 104

\bibitem{MNRAS-376-459}
  McCallum, J N, Ellingsen S P, Lovell J E J,
2007 {\it Mon.\ Not.\ Roy.\ Astron.\ Soc.}\
  {\bf 376} 549

\bibitem{ApJ-566-L9}
  Zavala R T and Taylor G B,
2002 {\it Astrophys.\ J.}\ {\bf
  566} L9

\bibitem{SovAL-6-42}
 Matveenko L I
 {\it et al.},
1980, {\it Sov.\ Astron.\ Lett.}\ {\bf 6} 42

\bibitem{AstronRep-51-808}
Artyukh V S and Chernikov P A,
2007 {\it Astron.\ Rep.}\ {\bf 51} 808

\bibitem{AstronRep-49-967}
Tyul'bashev S A,
2005,
{\it Astron.\ Rep.}\ {\bf 49} 967

\bibitem{AstronRep-50-202}
Chernikov P A
 {\it et al.},
2006 {\it Astron.\ Rep.}\ {\bf 50} 202

\bibitem{Slysh}
Slish V I,
1963
{\it Nature} {\bf 199} 682

\bibitem{Lukash}
Zakharov A F
  {\it et al.},
2003
  {\it Mon.\ Not.\ Roy.\ Astron.\ Soc.}\  {\bf 342} 1325

\bibitem{Gnedin}
  Gnedin Y M, Natsvlishvili T M and Piotrovich M Y,
  2005 {\it Grav.\ Cosmol.}\  {\bf 11} 333

\bibitem{Znajek}
  Znajek R L,
1978
{\it Mon.\ Not.\ Roy.\ Astron.\ Soc.}\ {\bf 185} 833

\bibitem{Abram}
Ghosh P and Abramowicz M A,
1997
{\it Mon.\ Not.\ Roy.\ Astron.\ Soc.}\ {\bf 292} 887

\bibitem{ShakuraSyunyaev}
  Shakura N I and Syunyaev R A,
 1973    {\it   Astron.\ Astrophys.}\  {\bf 24} 337

\bibitem{NovikovThorne}
Novikov I D and Thorne K S,
1973,
  in: {\it Black holes (Les astres occlus)}, Gordon and Breach,  p. 343

\bibitem{M-B_BH}
Zhang W M, Lu Y and Zhang S N,
2005
  {\it Chin.\ J.\ Astron.\ Astrophys.\ Suppl.}\ {\bf 5} 347

\bibitem{LEDA}
Paturel G  {\it et al.},
2003
{\it  Astron.\ Astrophys.}
  {\bf 412} 45;
http://leda.univ-lyon1.fr

\bibitem{NED}
The NASA/IPAC Extragalactic database (available at
http://nedwww.ipac.caltech.edu).

\bibitem{0607228}
  Harris D E and Krawczynski H,
2006
  {\it Ann.\ Rev.\ Astron.\ Astrophys.}\  {\bf 44} 463

\bibitem{0604219}
  Gallimore J F
  {\it et al.},
2006
  {\it Astron.\ J.}\  {\bf 132} 546

\bibitem{0204361}
Schurch N J, Roberts T P and Warwick R S,
2002
  {\it Mon.\ Not.\ Roy.\ Astron.\ Soc.}\  {\bf 335} 241

\bibitem{0510661}
Kataoka J {\it et al.},
2006
  {\it Astrophys.\ J.}\  {\bf 641} 158

\bibitem{0710.1277}
Hardcastle M J {\it et al.},
2007
 {\it Astrophys.\ J.}\  {\bf 670} L81

\bibitem{ApJ-511-686}
  Allen, M G
  {\it et al.},
1999 {\it Astrophys.\ J.}\ {\bf 511} 686

\bibitem{0709.4476}
  Laine S and Beck R,
  2008 {\it Astrophys.\ J.}\ {\bf 673} 128

\bibitem{ApJ-273-128}
Burns J O, Feigelson E D, Schreier E J,
1983 {\it Astrophys.\ J.}\ {\bf 273} 128

\bibitem{0306317}
  Schwartz D A
  {\it et al.},
  2003 {\it New Astron.\ Rev.}\  {\bf 47} 461

\bibitem{ApJ-612-729}
Hardcastle M J, Harris D E and Worrall D M,
2004 {\it Astrophys.\ J.}\  {\bf 612} 729

\bibitem{0904.3925}
Harris D E {\it et al.},
  arXiv:0904.3925

\bibitem{KataokaStawarz}
  Kataoka, J and Stawarz, L,
   {\it Astrophys.\ J.}\ {\bf 622} 797

\bibitem{Meisenheimer}
  Meisenheimer K
  {\it et al.},
        1989 {\it Astron.\ Astrophys.}\ {\bf 219} 63

\bibitem{HardcastleFRII}
  Croston J H
  {\it et al.},
2005 {\it Astrophys.\ J.}\ {\bf 626} 733
  [arXiv:astro-ph/0503203].

\bibitem{0510452}
  Vlemmings W H T
  {\it et al.},
  arXiv:astro-ph/0510452.

\bibitem{0605741}
  Slysh V I and Migenes V,
2006  {\it Mon.\ Not.\ Roy.\ Astron.\ Soc.}\  {\bf 369} 1497

\bibitem{0804.1141}
  Vlemmings W H T,
2008 {\it Astron.\ Astrophys.} {\bf 484} 773

\bibitem{ApJ-674-295}
  Sarma A P
  {\it et al.},
2008 {\it Astrophys.\ J.}\ {\bf 674} 295

\bibitem{0709.0256}
  Curran R L and Chrysostomou A,
2007
  {\it Mon.\ Not.\ Roy.\ Astron.\ Soc.}\  {\bf 382} 699

\bibitem{9903145}
Anchordoqui L A, Romero G E and Combi J A,
1999
  Phys.\ Rev.\  D {\bf 60} 103001

\bibitem{WaxmanSB}
  Thompson T A
  {\it et al.},
2006 {\it
  Astrophys.\ J.}\  {\bf 645} 186

\bibitem{GRB:Piran}
Piran T,
2005
  {\it AIP Conf.\ Proc.}\  {\bf 784} 164

\bibitem{0409489}
Lyutikov M,
  {\it Magnetic fields in GRBs},
  arXiv:astro-ph/0409489.

\bibitem{GRBGamma}
Piran T,
1999
  {\it Phys.\ Rept.}\  {\bf 314} 575

\bibitem{0410182}
  Govoni F and Feretti L,
2004
  {\it Int.\ J.\ Mod.\ Phys.}\  D {\bf 13} 1549

\bibitem{0801.0985}
  Ferrari C
  {\it et al.},
2008 {\it Space Science Reviews}, {\bf 134} 93

\bibitem{0507367}
Beck R and Krause M,
2005
{\it Astron.\ Nachr.}\  {\bf 326} 414

\bibitem{0704.3288}
Kronberg P P, Kothes R, Salter C J and Perillat P,
2007
{\it  Astrophys.\ J.}\  {\bf 659}, 267

\bibitem{0011281}
Clarke T E, Kronberg P P and Boehringer H,
2001
{\it Astrophys.\ J.}\  {\bf 547} L111

\bibitem{0602622}
Taylor G B {\it et al.},
2006
{\it Mon.\ Not.\ Roy.\ Astron.\ Soc.}\  {\bf 368} 1500

\bibitem{0608433}
Govoni F {\it et al.},
2006
{\it Astron.\ Astrophys.}\ {\bf 460} 425

\bibitem{DolagMF}
Dolag K
{\it et al.},
2005
  JCAP {\bf 0501} 009

\bibitem{SiglMF}
Sigl G, Miniati F and Ensslin T A,
2004
  Nucl.\ Phys.\ Proc.\ Suppl.\  {\bf 136} 224

\bibitem{Ginz}
Ginzburg V L, Syrovatskii S I,
{\it The origin of cosmic rays}, 1964, London: Pergamon Press

\end{thebibliography}
\end{document}